\documentclass{article}

\usepackage{PRIMEarxiv}

\usepackage[utf8]{inputenc} 
\usepackage[T1]{fontenc}    
\usepackage{hyperref}       
\usepackage{url}   
\usepackage{acro}
\DeclareAcronym{ecc}{
short = ECC,
long = error correcting code
}
\DeclareAcronym{ber}{
short = BER,
long  = bit error rate
}
\DeclareAcronym{asic}{
short        = ASIC,
long         = application-specific integrated-circuit
}
\DeclareAcronym{fpga}{
short = FPGA,
long  = field-programmable gate array
}
\DeclareAcronym{gpu}{
short = GPU,
long  = graphics processing unit
}
\DeclareAcronym{ldpc}{
short = LDPC,
long = low-density parity-check
}
\DeclareAcronym{nbldpc}{
short = NB-LDPC,
long = non-binary low-density parity-check
}
\DeclareAcronym{dvbs2}{
short = DVB-S2,
long = ETSI digital video broadcasting \nth{2} generation
}
\DeclareAcronym{cn}{
short = CN,
long = check node
}
\DeclareAcronym{vn}{
short = VN,
long = variable node
}
\DeclareAcronym{spa}{
short = SPA,
long = sum-product algorithm
}
\DeclareAcronym{emsa}{
short = EMSA,
long = extended min-sum algorithm
}
\DeclareAcronym{mma}{
short = MMA,
long = min-max algorithm
}
\DeclareAcronym{pcm}{
short = PCM,
long = parity-check matrix
}
\DeclareAcronym{bn}{
short = BN,
long = bit node
}
\DeclareAcronym{qcldpc}{
short = QC-LDPC,
long = quasi-cyclic low-density parity-check
}
\DeclareAcronym{gf}{
short = GF,
long = Galois field
}
\DeclareAcronym{mlgd}{
short = MLDG,
long =  majority-logic decoding
}
\DeclareAcronym{ihrb}{
short = IHRB,
long =  iterative hard reliability-based
}
\DeclareAcronym{isrb}{
short = ISRB,
long =  iterative soft reliability-based
}
\DeclareAcronym{gbfda}{
short = GBFDA,
long =  generalized bit-flipping decoding algorithm
}
\DeclareAcronym{amsa}{
short = AMSA,
long =  Adaptive Multiset Stochastic Algorithm
}
\DeclareAcronym{wbrb}{
short = WBRB,
long =  weighted bit reliability-based
}
\DeclareAcronym{fbrb}{
short = FBRB,
long =  full bit reliability-based
}
\DeclareAcronym{mimo}{
short = MIMO,
long =  multiple-input multiple-output
}
\DeclareAcronym{qam}{
short = QAM,
long =  quadrature amplitude modulation
}
\DeclareAcronym{snr}{
short = SNR,
long =  signal-to-noise ratio
}
\DeclareAcronym{adbp}{
short = ADBP,
long =  analog digital belief propagation
}
\DeclareAcronym{srb}{
short = SRB,
long =  symbol reliability based
}
\DeclareAcronym{gps}{
short = GPS,
long =  global positioning system
}
\DeclareAcronym{vlsi}{
short = VLSI,
long =  very large scale integration
}
\DeclareAcronym{fht}{
short = FHT,
long =  fast Hadamard transform
}
\DeclareAcronym{fft}{
short = FFT,
long =  fast Fourier transform
}
\DeclareAcronym{hls}{
short = HLS,
long =  high-level synthesis
}
\DeclareAcronym{rtl}{
short = RTL,
long =  register transfer level
}

\DeclareAcronym{qos}{
short = QoS,
long =  quality of service
}

\DeclareAcronym{bp}{
short = BP,
long =  belief propagation
}

\DeclareAcronym{llr}{
short = LLR,
long =  log-likelihood ratio
}

\DeclareAcronym{csr}{
short = CSR,
long =  compressed sparse row
}

\DeclareAcronym{csc}{
short = CSC,
long =  compressed sparse column
}

\DeclareAcronym{lut}{
short = LUT,
long =  look-up table
}

\DeclareAcronym{cpu}{
short = CPU,
long  = central processing unit
}

\DeclareAcronym{sm}{
short = SM,
long  = streaming multiprocessor
}

\DeclareAcronym{ram}{
short = RAM,
long  = random access memory
}

\DeclareAcronym{ccsds}{
short = CCSDS,
long  = Consultative Committee for Space Data Systems
}

\DeclareAcronym{bch}{
short = BCH,
long  = Bose–Chaudhuri–Hocquenghem
}

\DeclareAcronym{fwht}{
short = FWHT,
long  = fast Walsh-Hadamard transform
}

\DeclareAcronym{tdp}{
short = TDP,
long  = thermal design power
}

\DeclareAcronym{cuda}{
short = CUDA,
long  = compute unified device architecture
}

\DeclareAcronym{pim}{
short = PiM,
long  = processing-in-memory
}

\DeclareAcronym{pnm}{
short = PnM,
long  = processing-near-memory
}

\DeclareAcronym{pum}{
short = PuM,
long  = processing-using-memory
}

\DeclareAcronym{dpu}{
short = DPU,
long  = DRAM processing unit
}

\DeclareAcronym{msa}{
short = MSA,
long = min-sum algorithm
}

\DeclareAcronym{qldpc}{
short = QLDPC,
long = quantum low-density parity-check
}

\DeclareAcronym{soc}{
short = SoC,
long = system on a chip
}

\DeclareAcronym{qec}{
short = QEC,
long = quantum error correction
}

\DeclareAcronym{css}{
short = CSS,
long = Calderbank-Shor-Steane
}

\DeclareAcronym{alu}{
short = ALU,
long = arithmetic logic unit
}

\usepackage{booktabs}       
\usepackage{amsmath,amssymb,amsfonts}
\usepackage{nicefrac}       
\usepackage{microtype}      
\usepackage{lipsum}
\usepackage{fancyhdr}       
\usepackage{graphicx}       
\graphicspath{{media/}}     
\usepackage{algorithmic}
\usepackage{algorithm}
\usepackage{comment}
\usepackage{multirow}
\usepackage{xcolor}
\usepackage{soul}
\usepackage{bm}
\usepackage{tcolorbox}
\usepackage{siunitx}
\usepackage{subcaption}

\title{GPU-Accelerated Syndrome Decoding for Quantum LDPC Codes below the 63 µs Latency Threshold
}

\author{
  Oscar Ferraz$^{1,2}$, Bruno Coutinho$^{2}$, Gabriel Falcao$^{1,2}$, Marco Gomes$^{1, 2}$,\\ \textbf{Francisco A. Monteiro$^{2,3}$, Vitor Silva$^{1,2}$}\\
$^{1}$Department of Electrical and Computer Engineering, University of
Coimbra, Coimbra, Portugal\\
$^{2}$Instituto de Telecomunicações, Portugal\\
$^{3}$ISCTE -- Instituto Universitário de Lisboa, Lisbon, Portugal\\
  \texttt{\{oscar.ferraz, gff, marco, vitor\}@co.it.pt;}\\ \texttt{\{bruno.coutinho, francisco.monteiro\}@lx.it.pt}
}

\begin{document}
\maketitle

\begin{abstract}
This paper presents a GPU-accelerated decoder for quantum low-density parity-check (QLDPC) codes that achieves sub-$63$ $\mu$s latency, below the surface code decoder's real-time threshold demonstrated on Google’s Willow quantum processor. While surface codes have demonstrated below-threshold performance, the encoding rates approach zero as code distances increase, posing challenges for scalability. Recently proposed QLDPC codes, such as those by Panteleev and Kalachev, offer constant-rate encoding and asymptotic goodness but introduce higher decoding complexity. To address such limitation, this work presents a parallelized belief propagation decoder leveraging syndrome information on commodity GPU hardware. Parallelism was exploited to maximize performance within the limits of target latency, allowing decoding latencies under $50$ $\mu$s for [[$784$, $24$, $24$]] codes and as low as  $23.3$ $\mu$s for smaller codes, meeting the tight timing constraints of superconducting qubit cycles. These results show that real-time, scalable decoding of asymptotically good quantum codes is achievable using widely available commodity hardware, advancing the feasibility of fault-tolerant quantum computation beyond surface codes.
\end{abstract}


\section{Introduction}

\Ac{qec} is a fundamental requirement for building of fault-tolerant quantum computers due to the delicate nature of quantum information and the relatively high noise rates in physical quantum devices~\cite{nielsen:2010, acharya2024quantum, hanzo2025quantum, cruz2023quantum}. Superconducting qubits, such as those used in Google’s Willow processor, experience error rates in the order of $10^{-3}$ to $10^{-2}$ per gate~\cite{ai2023suppressing} (without \ac{qec}), far above what is tolerable without active error correction~\cite{acharya2024quantum}. In $2024$, Google demonstrated the first instance of fault-tolerant quantum computation using a distance-$5$ surface code and achieved an average decoder latency of $63$ µs for real-time syndrome processing over a million cycles~\cite{acharya2024quantum}. 


However, while surface codes are well-suited to current hardware due to their locality and high thresholds, the encoding rate (ratio of logical to physical qubits) asymptotically approaches zero as the code distance increases~\cite{devitt2013quantum}, making them suboptimal for large-scale quantum computation. This leads to substantial physical qubit overheads in large-scale systems.

A promising alternative lies in \ac{qldpc} codes such as those by Panteleev and Kalachev~\cite{panteleev2021quantum}, which are “good” in the information-theoretic sense, maintaining constant encoding rates and linear minimum distance scaling~\cite{Coffey:1990}. Moreover, recent constructions~\cite{dinur2023good} prove the existence of \ac{qldpc} codes with linear-time decoders, opening new avenues for practical and scalable quantum error correction.

Yet, the decoding of \ac{qldpc} codes is more intricate than that of surface codes. Their larger block lengths, sparse parity-check matrices, and often nonlocal stabilizers impose significant computational burdens on classical decoding hardware when aiming to meet sub-$63$ $\mu$s latency targets~\cite{acharya2024quantum}.


A key distinction explored in this work lies in the fundamental difference between classical and quantum \ac{ldpc} decoding. Classical \ac{ldpc} decoding algorithms assume access to the noisy codeword, as is typical in communication systems~\cite{ren2024generalized, ferraz2021survey}, allowing direct use of soft information (such as \acp{llr}) to iteratively update messages in a factor graph. In contrast, \ac{qec} must operate under the constraint that directly measuring the quantum state collapses its superposition, thereby destroying the encoded information.

To circumvent this, \ac{qldpc} decoding relies on syndrome-based decoding~\cite{valls2021syndrome, yao2024belief}, where stabilizer measurements are used to extract error syndromes, with the help of ancilla qubits, without disturbing the underlying quantum state. This enables error detection and correction while preserving quantum coherence.

Unlike classical decoders, which can observe and process the entire corrupted codeword, quantum decoders must infer the most likely error pattern solely from the measured syndrome, which indicates which stabilizer constraints have been violated. This imposes a fundamentally more challenging inference problem and alters the decoding dynamics~\cite{raveendran2023soft}. Moreover, the syndrome extraction process in quantum hardware is itself subject to noise, further complicating the decoding task and motivating the use of soft-syndrome information to support probabilistic reasoning and improve convergence under uncertainty~\cite{yao2024belief}.





This work presents a \ac{gpu}-accelerated, belief-propagation-based decoder for \ac{qldpc} codes that leverages syndrome inputs and achieves latencies below the $63$~$\mu$s threshold. Implemented on \ac{gpu} platforms, the decoder exploits the fine-grained parallelism, shared memory, and reduced-precision arithmetic features of desktop and \ac{soc} \acp{gpu}. Kernel optimizations and warp-aware memory layouts were designed to minimize thread divergence and maximize throughput. The decoder implementation is based on codes structurally similar to~\cite{dinur2023good, panteleev2021quantum, bravyi2024high}. While different quantum hardware platforms may impose distinct timing constraints, this work specifically targets the $63$~$\mu$s latency threshold demonstrated in Google's Willow processor~\cite{acharya2024quantum}, with the goal of providing a viable solution for real-time fault-tolerant quantum error correction under such conditions.

This paper proposes the following contributions:
\begin{itemize}
  \item The first \ac{gpu}-based \ac{qldpc} decoder implementation, to the best of our knowledge.
  \item An efficient mapping of the \ac{qldpc} decoder to minimize latency and maximize throughput in \ac{gpu} systems, exploiting the architectures' advantages and avoiding its drawbacks and limitations.
\end{itemize}

\section{GPU-based QLDPC Decoder}

The syndrome-based \ac{ldpc} decoding process does not rely on direct access to the transmitted codeword, as in classical communication systems, but instead infers the most likely error configuration based solely on the syndrome vector. This is particularly relevant in the context of quantum \ac{ldpc} decoding, where the syndrome, extracted via stabilizer measurements, provides indirect information about which qubits were affected by errors~\cite{breuckmann2021quantum}.

\begin{figure}[t!]
\begin{center}
\includegraphics[width=0.8\columnwidth]{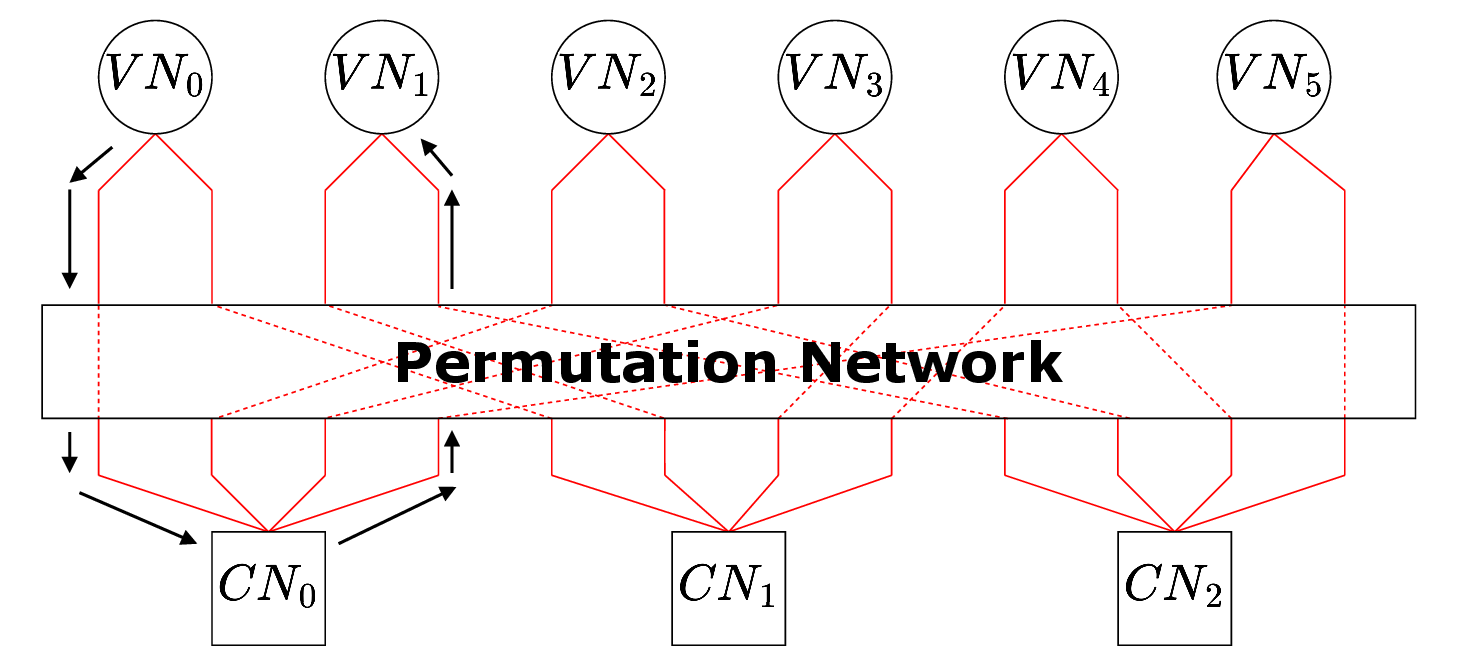}
\caption{\small \ac{ldpc} Tanner graph based on (\ref{eq:PCM}). The number of rows of $\mathbf{H}$ corresponds to the number of \acp{cn} ($M$), while the number of columns represents the number of \acp{vn} ($N$). The non-zero elements of $\mathbf{H}$ dictate the connections between \acp{vn} and \acp{cn}. After initialization, the messages are propagated from \acp{vn} to the connected \acp{cn} and back in an iterative manner until the parity-check equation is verified or until a maximum number of iterations has been reached, as described in Algorithm~\ref{alg:MS}~\cite{raveendran2023soft}. }
\label{fig:Tanner_graph}
\end{center}
\end{figure}

An \ac{ldpc} code is defined by a parity-check matrix $\mathbf{H}$, which specifies the structure of a bipartite Tanner graph composed of \acp{cn} and \acp{vn}. The number of \acp{cn} and \acp{vn} are given by $M$ and $N$, respectively, and their interconnections are determined by the non-zero entries in 
$\mathbf{H}$. For example, the following matrix
%
\begin{equation} \label{eq:PCM}
\mathbf{H}= \begin{bmatrix}
1  & 0  &1  &1 &0  &1 \\ 
1 & 1   &0  &1  &1  &0 \\ 
0 &1   &1  &0  &1  &1 
\end{bmatrix},
\end{equation}
%
defines a simple \ac{ldpc} code with six \acp{vn} and three \acp{cn}. The sets $S_v(m)$ and $S_c(n)$ identify which \acp{vn} are connected to each \ac{cn}, and vice versa. This code has a \ac{cn} degree of four ($d_c=4$) and a \ac{vn} degree of two ($d_v=2$), meaning each \ac{cn} is connected to four \acp{vn} and each \ac{vn} to two \acp{cn}. The associated Tanner graph is illustrated in Fig.~\ref{fig:Tanner_graph}.

In syndrome-based decoding, the decoder has no access to the original transmitted codeword or the channel likelihoods. Instead, it receives a syndrome vector $\mathbf{s}=\mathbf{H}\cdot \mathbf{e^T}$, where $\mathbf{e}$ is the unknown error pattern affecting the true codeword, and all the calculations are assumed to be carried out in $GF(2)$~\cite{carrasco2008non, moon2020error}. The goal is to find the most probable $\mathbf{e}$ such that the observed syndrome is satisfied, i.e.,  $\mathbf{H}\cdot \mathbf{e^T}=\mathbf{s}$ 

In classical channel-based \ac{ldpc} decoding, the decoder typically operates on a received word  $\mathbf{y}=\mathbf{c}+\mathbf{e}$, where $\mathbf{c}$ is the transmitted codeword. The decoder uses channel \acp{llr} derived from $\mathbf{y}$ to estimate $\mathbf{c}$. The syndrome can also be computed directly from $\mathbf{y}$ as $\mathbf{s}=\mathbf{H}\cdot \mathbf{y^T}$, since $\mathbf{H}\cdot \mathbf{c^T}=\mathbf{0}$ by definition of $\mathbf{c}$ being a codeword. This yields the same result as $\mathbf{s}=\mathbf{H}\cdot( \mathbf{c}+\mathbf{e})^T=\mathbf{H}\cdot \mathbf{e}^T$ ($\mathbf{s}$, $\mathbf{e}$, $\mathbf{y}$ and $\mathbf{c}$ are defined as row-vectors).

\begin{algorithm}[t!]
\caption{MSA for syndrome-based LDPC decoding}
\label{alg:MS}
{\footnotesize
\begin{algorithmic}[1]
\STATE {\bf Input:} syndrome ($\mathbf{s}$), parity-check matrix ($\mathbf{H}$), max. number of iterations ($\mathbf{I_{max}}$) \\
\STATE {\bf Initialization:} $q^{(0)}_{m,n} = \gamma_n = 1$, $k = 1$ \\
\REPEAT
    \STATE {\bf Check node processing:}
    \FOR{each check node $m$, each $n \in S_v(m)$}
        \STATE $r^{(k)}_{m,n}= \left (\alpha s_m\prod\limits_{j\in S_v(m)\backslash n}\left ( \operatorname{sign}\left ( q^{(k-1)}_{m,j} \right ) \right ) \right )\min\limits_{j\in S_v(m)\backslash n}\left ( \left | q^{(k-1)}_{m,j} \right | \right )$
    \ENDFOR
    \STATE {\bf Variable node processing:}
    \FOR{each variable node $n$, each $m \in S_c(n)$}
        \STATE $q^{(k)}_{m,n} = \gamma_n + \sum\limits_{i \in S_c(n) \backslash m} r^{(k)}_{i,n}$
    \ENDFOR
    \STATE {\bf A posteriori calculation \& Tentative decision:}
    \FOR{each variable node $n$}
        \STATE $Q^{(k)}_{n} = \gamma_n + \sum\limits_{i \in S_c(n)} r^{(k)}_{i,n}$
        \STATE $e^{(k)}_{n} = \begin{cases} 
            1 & \text{if } Q^{(k)}_{n} < 0 \\
            0 & \text{if } Q^{(k)}_{n} \geq 0
        \end{cases}$
    \ENDFOR
\UNTIL{$k < I_{max}$ \text{ and } $\left(\mathbf{e^{(k-1)}} \mathbf{H}^T \right) \neq \mathbf{s}$}
\end{algorithmic}
}
\end{algorithm}

However, in \ac{qldpc} decoding, the actual received word $\mathbf{y}$ is not observed, since the decoder only has access to the syndrome $\mathbf{s}$. Therefore, decoding must proceed without channel likelihoods. In this setting, soft-syndrome decoding extends traditional binary syndrome decoding by associating analog reliability values to each syndrome bit~\cite{raveendran2023soft}. These can be derived from repeated quantum measurements, circuit-level noise simulations, or confidence estimates in the extraction circuit~\cite{pecorari2025high}. The decoder then uses this soft information to prioritize more likely error patterns and accelerate convergence.

\begin{figure}[t!]
\begin{center}
\includegraphics[width=0.4\columnwidth]{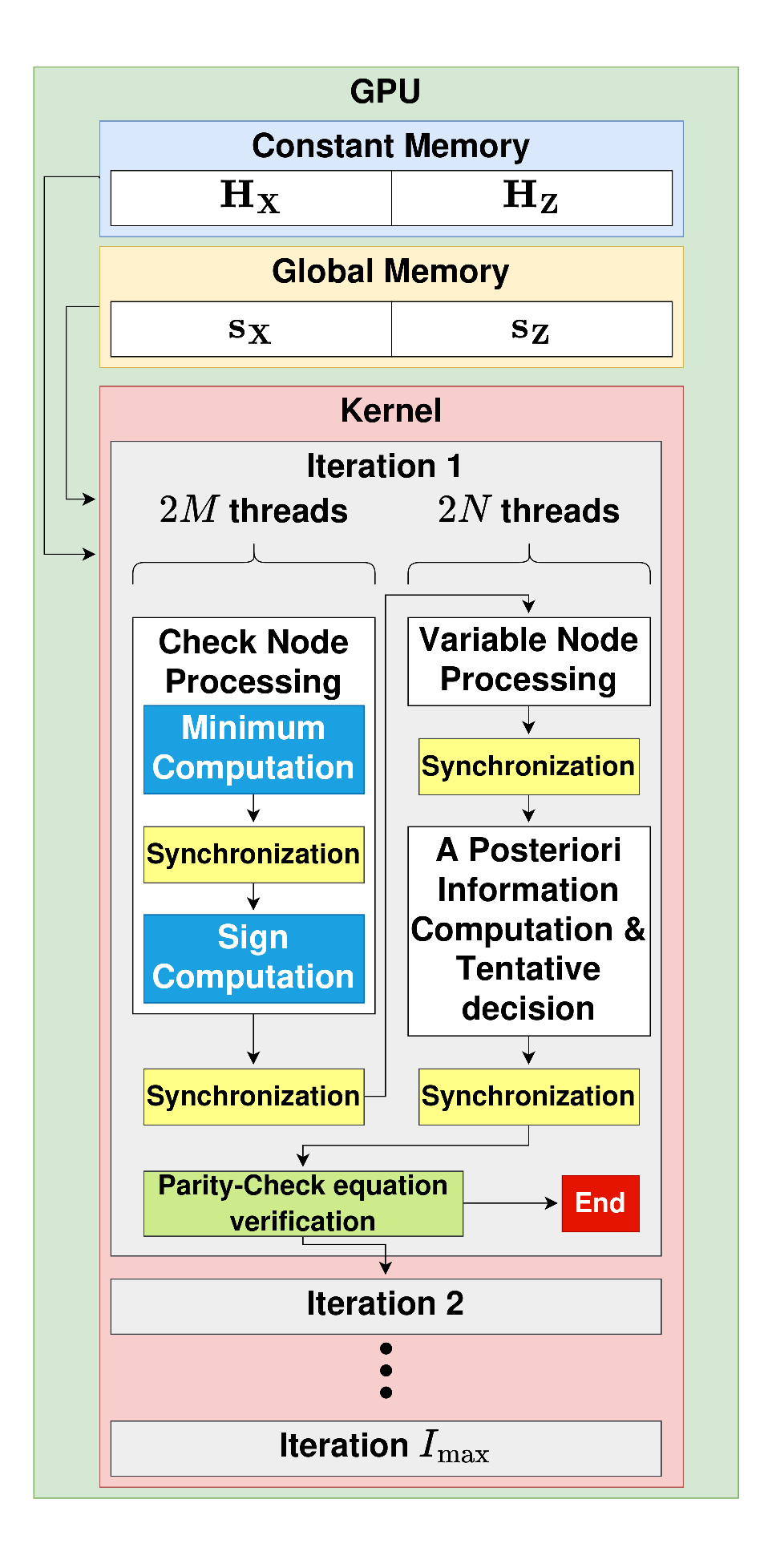}
\caption{\small Proposed \ac{gpu}-based \ac{qldpc} decoder. The \ac{qldpc} parity-check matrices $\mathbf{H_X}$ and $\mathbf{H_Z}$ and stored in the compressed format in constant memory for faster accesss. The remaining data, such as the syndromes  $\mathbf{s_X}$ and $\mathbf{s_Z}$, is stored in global memory. Each decoding iteration has four synchronization barriers and the \ac{cn} processing is further divided into the minimum and sign computations. The \ac{cn} processing is executed with $2M$ threads (equal to the size of  $\mathbf{s_X}+\mathbf{s_Z}$) while the \ac{vn} processing, a posteriori information, and tentative decision are processed with $2N$ threads. Both X and Z components are concatenated in memory and decoded in the same kernel, reducing latency and increasing throughput.}
\label{fig:LDPC}
\end{center}
\end{figure}

\begin{figure*}[!t]
\begin{center}
\includegraphics[width=0.75\textwidth]{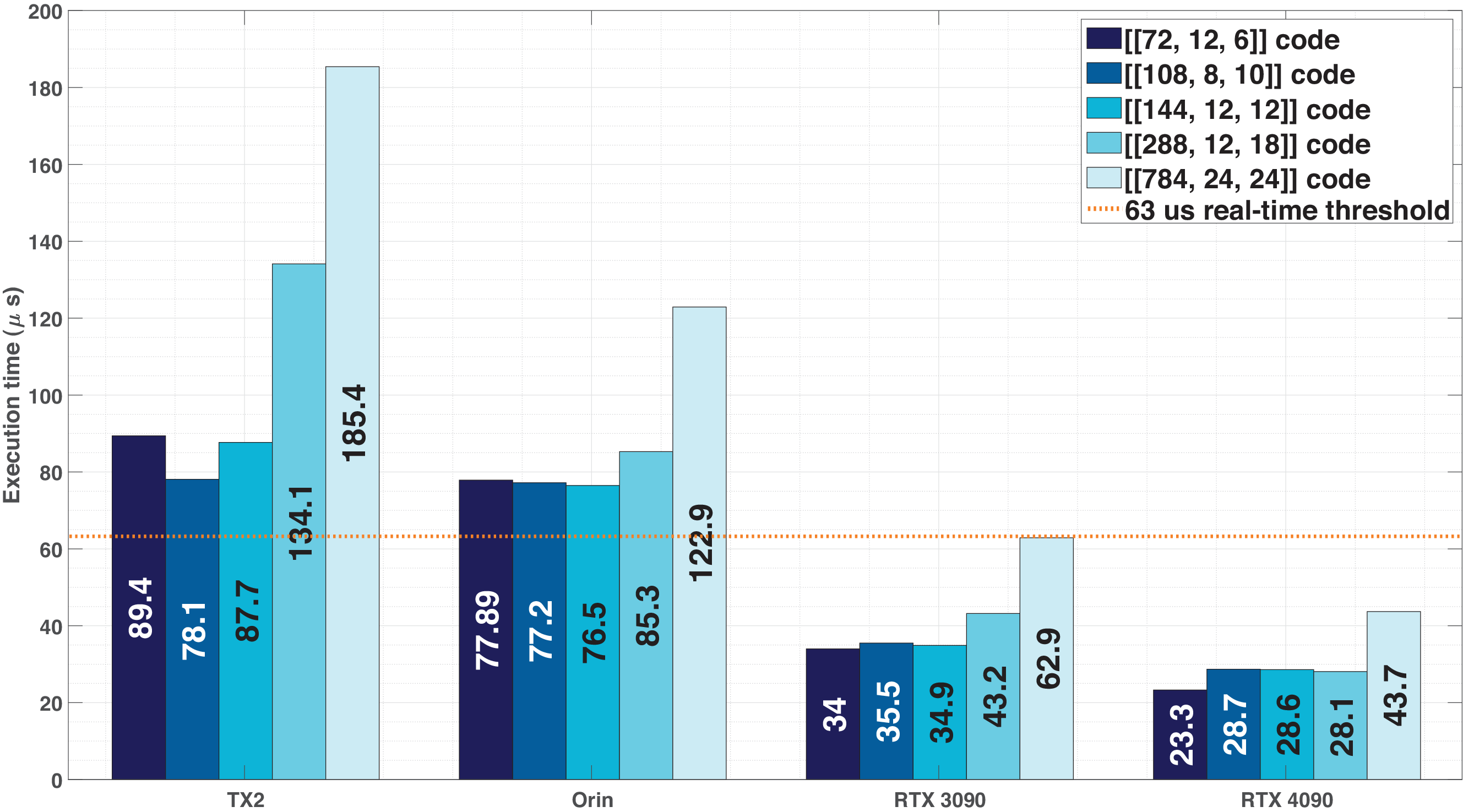}
\caption{\small Decoding performance for the \ac{qldpc} codes [[$72$, $12$, $6$]], [[$108$, $8$, $10$]], [[$144$, $12$, $12$]], [[$288$, $12$, $18$]], and [[$784$, $24$, $24$]] provided in~\cite{bravyi2024high}. The bars represent the decoding time considering data transfers between host-to-device and device-to-host and kernel execution time for ten decoding iteration using floating-point arithmetic precision in several \acp{gpu}. The dashed line represent the $63$ $\mu$s real-time threshold~\cite{acharya2024quantum}.}
\label{fig:chart}
\end{center}
\end{figure*}

In \ac{qec}, particularly under the stabilizer formalism, \ac{qldpc} codes are constructed by combining two classical binary \ac{ldpc} codes to separately detect and correct \emph{bit-flip} (X) and \emph{phase-flip} (Z) errors~\cite{babar2015fifteen}. These two types of errors are associated with the non-commuting Pauli operators and must be corrected independently to preserve quantum information. The codes are defined by two parity-check matrices, $\mathbf{H_X} \in GF(2)$ and $\mathbf{H_Z} \in GF(2)$, where each row of $\mathbf{H_X}$ (respectively, $\mathbf{H_Z}$) corresponds to an X-type (respectively, Z-type) stabilizer generator acting on the $n$ physical qubits~\cite{devitt2013quantum}.

To define a valid \ac{css}-type quantum code~\cite{rengaswamy2020optimality, calderbank1996good}, the stabilizer generators must commute, which imposes the commutativity condition $\mathbf{H_X}\mathbf{H_Z^T} = \mathbf{0} $, which ensures that the X and Z stabilizer generators commute, a necessary property for the stabilizer group to form a valid quantum code space. This condition guarantees that the simultaneous eigenstates of all stabilizers define a proper code subspace.

During quantum error correction, syndromes are measured independently for X and Z errors using their respective stabilizers $\mathbf{s_Z} = \mathbf{H_X} \mathbf{e_Z}^T $ and $\mathbf{s_X} = \mathbf{H_Z} \mathbf{e_X}^T $, where $\mathbf{e_X}$, $\mathbf{e_Z}$ are binary vectors representing the X and Z error patterns, respectively. Two independent syndrome decoders are then used to estimate error vectors $\mathbf{\hat{e}_X}$ and $\mathbf{\hat{e}_Z}$ such that $\mathbf{H_Z} \mathbf{\hat{e}_X}^T=\mathbf{s_X}$ and $\mathbf{H_X} \mathbf{\hat{e}_Z}^T=\mathbf{s_Z}$.

To perform this decoding in practice, classical \ac{ldpc} decoding algorithms can be applied separately to the $\mathbf{s_X}$ and $\mathbf{s_Z}$ syndrome equations, treating each as an independent binary decoding problem over $GF(2)$. Given the sparse nature of the parity-check matrices $\mathbf{H_X}$ and $\mathbf{H_Z}$, iterative message-passing algorithms are well-suited for this task, enabling efficient approximation of the most likely error patterns. 

Among the various \ac{ldpc} decoding algorithms, this work adopts the \ac{msa}~\cite{fossorier1999reduced} for its low complexity and good performance in sparse-graph decoding~\cite{ferraz2023unlocking}. Algorithm~\ref{alg:MS}~\cite{raveendran2023soft} outlines the decoding procedure, where each \ac{cn} computes a new message based on incoming \ac{vn} messages, and each \ac{vn} aggregates information to update its belief of certainty of a given bit. This algorithm also features a scaling factor $\alpha \in (0,1)$ in the \ac{cn} processing step to mitigate the overestimation of message magnitudes and improve decoding accuracy compared to the standard formulation.

The proposed \ac{gpu}-based \ac{qldpc} decoder implementation uses a single kernel to execute both X and Z components. As illustrated in Fig.~\ref{fig:LDPC}, each decoding iteration is executed within one \ac{gpu} kernel and comprises multiple computational stages mapped onto distinct threads. The process begins with the \textit{\ac{cn} processing}, which includes the computation of message signs and magnitudes, followed by \textit{\ac{vn} processing}, \textit{a posteriori information computation}, and \textit{tentative decision}. Constant memory allows to exploit faster read times and is used to store parity-check matrices $\mathbf{H_X}$ and $\mathbf{H_Z}$ which are static structures. After each \ac{cn} and \ac{vn} update (and inside the \ac{cn} processing), synchronization barriers ensure consistent memory states across threads before proceeding to the next operation or iteration. The decoding loop proceeds until a stopping criterion is met (e.g., maximum number of iterations or valid syndrome check). This architecture exploits fine-grained parallelism by assigning individual \ac{cn} and \ac{vn} operations to CUDA threads, achieving high throughput.

\begin{table*}[!t]
    \begin{center}
        \caption{\small Topics to be discussed in the full paper.}
        \label{table:specs}
        \resizebox{\textwidth}{!}{ 
            \begin{tabular}{|l|l|}
            \hline
             \multicolumn{1}{|c|}{\textbf{Topic}} & \multicolumn{1}{c|}{\textbf{Issues and Optimization Guidelines}} \\ \hline \hline
            Different \ac{qldpc} codes & Different \ac{qldpc} codes can provide better error-correction capability at low latencies. \\ \hline
            Float vs. Integer Precision & Integer arithmetic offers higher performance and lower memory usage, at the cost of using quantization schemes to preserve decoding accuracy. \\ \hline
            Reduced Precision Data Types & 8- and 16-bit integer formats can reduce data transfer volume by 4× and 2×, respectively, contributing to lower latency and improved throughput. \\ \hline
            Thread and Block Configuration & Optimizing thread and block dimensions can increase kernel occupancy, reduce divergence, and improve memory coalescing. \\ \hline
            Shared vs. Global Memory & Using shared memory for frequently accessed data reduces memory access latency and improves throughput compared to global memory access. \\ \hline
            Asynchronous Kernel Execution & Launching multiple decoders asynchronously allows overlapping data transfers with kernel execution, effectively hiding memory latency. \\ \hline
            Quantum Device Thresholds & Compare the proposed solution's latency with the timing constraints of different quantum devices. \\ \hline
            \end{tabular}
        }
    \end{center}
\end{table*}
   
\section{Preliminary Results}

The preliminary results use binary \ac{qldpc} codes [[$72$, $12$, $6$]], [[$108$, $8$, $10$]], [[$144$, $12$, $12$]], [[$288$, $12$, $18$]], and [[$784$, $24$, $24$]] (the code parameter [[$n$, $k$, $d$]] encodes $k$ logical qubits into $n$ data qubits offering a code distance $d$~\cite{bravyi2024high}), as provided in~\cite{bravyi2024high} running ten decoding iterations without early termination. In this extended abstract, the decoder is implemented using floating-point arithmetic precision and it is used constant memory for storing static structures (parity-check matrices) and global memory for the remaining data. The devices used are the Nvidia Jetson TX$2$, Jetson AGX Orin, RTX $3090$ and RTX $4090$.


Fig.~\ref{fig:chart} compares the average decoding latency across four devices using five \ac{qldpc} codes. On the Jetson platforms, the latency exceeds the  $63$ $\mu$s threshold by approximately 20~\(\mu\)s for the smaller codes, with larger codes exhibiting even more pronounced deviations. In contrast, on the RTX $3090$ and $4090$ \acp{gpu}, the implementation achieves sub-threshold latencies for all codes. Specifically, the [[$784$, $24$, $24$]] code reaches $62.9$ $\mu$s and $43.7$ $\mu$s on the $3090$ and $4090$, respectively, while the remaining codes achieve even lower latencies of $35$ $\mu$s and $28$ $\mu$s.

Profiling metrics from the implementations reveal that the decoding kernel accounts for the majority of the total latency. For smaller codes, data transfer between host and device (both host-to-device and device-to-host) contributes approximately $15$\% of the total latency, while for larger codes, this overhead is reduced to around $5$\%, as kernel execution becomes increasingly dominant. Although the Jetson platforms exhibit lower data transfer overheads, due to memory shared between host and device eliminating the need for explicit transfers, their reduced computational performance results in higher overall latency. In contrast, the more powerful desktop \acp{gpu} (RTX $3090$ and $4090$) are able to perform faster kernel execution, ultimately achieving lower end-to-end latencies.

While the results presented in this extended abstract demonstrate the feasibility of low-latency decoding for \ac{qldpc} codes on \ac{gpu} architectures, several optimization strategies remain to be explored to further reduce latency and improve throughput, as shown in Table~\ref{table:specs}. The adoption of lower-precision arithmetic: using $8$-bit or $16$-bit integer data types, combined with appropriate quantization schemes of the soft-syndromes, can significantly improve kernel performance and reduce memory usage. These representations also cut data transfer volumes by factors of $4\times$ and $2\times$, respectively, offering a practical way to reduce overall decoding latency. Furthermore, integer \acp{alu} can achieve better performance compared to floating-point \acp{alu}.

Additionally, fine-tuning the thread and block configuration, along with exploiting shared memory in place of global memory for frequently accessed variables, can yield further gains in computational efficiency. Shared memory access is significantly faster than global memory, and careful use of this resource can reduce memory access latency and increase overall throughput.

Another promising optimization is the use of asynchronous kernel launches, allowing multiple decoders to operate in parallel and overlap data transfers with kernel execution. This strategy effectively hides memory transfer latency behind computation and is particularly beneficial when decoding batches of syndromes or performing pipeline parallelism across multiple code blocks.


In addition to targeting the $63$~$\mu$s threshold demonstrated by Google's Willow processor, future work will also include a comparative analysis of our decoder's performance against the timing constraints imposed by other quantum computing platforms, which may allow for more relaxed or more stringent latency requirements depending on their gate times and error correction cycles.

These directions collectively offer a roadmap for achieving even faster and more scalable \ac{qldpc} decoders, bringing us closer to practical real-time quantum error correction in future fault-tolerant quantum computing systems.


\bibliographystyle{unsrt}  
\bibliography{references}

\begin{thebibliography}{10}

\bibitem{nielsen:2010}
Michael~A Nielsen and Isaac~L Chuang.
\newblock {\em {Quantum computation and quantum information}}.
\newblock Cambridge university press, 2010.

\bibitem{acharya2024quantum}
Rajeev Acharya, Dmitry~A Abanin, Laleh Aghababaie-Beni, Igor Aleiner, Trond~I Andersen, Markus Ansmann, Frank Arute, Kunal Arya, Abraham Asfaw, Nikita Astrakhantsev, et~al.
\newblock {Quantum error correction below the surface code threshold}.
\newblock {\em Nature}, 2024.

\bibitem{hanzo2025quantum}
Lajos Hanzo, Zunaira Babar, Zhenyu Cai, Daryus Chandra, Ivan~B Djordjevic, Balint Koczor, Soon~Xin Ng, Mohsen Razavi, and Osvaldo Simeone.
\newblock {Quantum Information Processing, Sensing, and Communications: Their Myths, Realities, and Futures}.
\newblock {\em Proceedings of the IEEE}, 2025.

\bibitem{cruz2023quantum}
Diogo Cruz, Francisco~A Monteiro, and Bruno~C Coutinho.
\newblock {Quantum error correction via noise guessing decoding}.
\newblock {\em IEEE Access}, 11:119446--119461, 2023.

\bibitem{ai2023suppressing}
Google~Quantum AI.
\newblock {Suppressing quantum errors by scaling a surface code logical qubit}.
\newblock {\em Nature}, 614(7949):676, 2023.

\bibitem{devitt2013quantum}
Simon~J Devitt, William~J Munro, and Kae Nemoto.
\newblock {Quantum error correction for beginners}.
\newblock {\em Reports on Progress in Physics}, 76(7):076001, 2013.

\bibitem{panteleev2021quantum}
Pavel Panteleev and Gleb Kalachev.
\newblock {Quantum LDPC codes with almost linear minimum distance}.
\newblock {\em IEEE Transactions on Information Theory}, 68(1):213--229, 2021.

\bibitem{Coffey:1990}
J.T. Coffey and R.M. Goodman.
\newblock {Any code of which we cannot think is good}.
\newblock {\em IEEE Transactions on Information Theory}, 36(6):1453--1461, 1990.

\bibitem{dinur2023good}
Irit Dinur, Min-Hsiu Hsieh, Ting-Chun Lin, and Thomas Vidick.
\newblock {Good quantum LDPC codes with linear time decoders}.
\newblock In {\em Proceedings of the 55th annual ACM symposium on theory of computing}, pages 905--918, 2023.

\bibitem{ren2024generalized}
Yuqing Ren, Hassan Harb, Yifei Shen, Alexios Balatsoukas-Stimming, and Andreas Burg.
\newblock {A generalized adjusted min-sum decoder for 5G LDPC codes: Algorithm and implementation}.
\newblock {\em IEEE Transactions on Circuits and Systems I: Regular Papers}, 2024.

\bibitem{ferraz2021survey}
Oscar Ferraz, Srinivasan Subramaniyan, Ramesh Chinthala, Jo{\~a}o Andrade, Joseph~R Cavallaro, Soumitra~K Nandy, Vitor Silva, Xinmiao Zhang, Madhura Purnaprajna, and Gabriel Falcao.
\newblock A survey on high-throughput non-binary ldpc decoders: Asic, fpga, and gpu architectures.
\newblock {\em IEEE Communications Surveys \& Tutorials}, 24(1):524--556, 2021.

\bibitem{valls2021syndrome}
Javier Valls, Francisco Garcia-Herrero, Nithin Raveendran, and Bane Vasi{\'c}.
\newblock {Syndrome-based min-sum vs OSD-0 decoders: FPGA implementation and analysis for quantum LDPC codes}.
\newblock {\em IEEE Access}, 9:138734--138743, 2021.

\bibitem{yao2024belief}
Hanwen Yao, Waleed~Abu Laban, Christian H{\"a}ger, Alexandre~Graell i~Amat, and Henry~D Pfister.
\newblock {Belief propagation decoding of quantum LDPC codes with guided decimation}.
\newblock In {\em 2024 IEEE International Symposium on Information Theory (ISIT)}, pages 2478--2483. IEEE, 2024.

\bibitem{raveendran2023soft}
Nithin Raveendran, Javier Valls, Asit~Kumar Pradhan, Narayanan Rengaswamy, Francisco Garcia-Herrero, and Bane Vasi{\'c}.
\newblock {Soft syndrome iterative decoding of quantum LDPC codes and hardware architectures}.
\newblock {\em EPJ Quantum Technology}, 10(1):45, 2023.

\bibitem{bravyi2024high}
Sergey Bravyi, Andrew~W Cross, Jay~M Gambetta, Dmitri Maslov, Patrick Rall, and Theodore~J Yoder.
\newblock {High-threshold and low-overhead fault-tolerant quantum memory}.
\newblock {\em Nature}, 627(8005):778--782, 2024.

\bibitem{breuckmann2021quantum}
Nikolas~P Breuckmann and Jens~Niklas Eberhardt.
\newblock {Quantum low-density parity-check codes}.
\newblock {\em PRX Quantum}, 2(4):040101, 2021.

\bibitem{carrasco2008non}
Rolando~Antonio Carrasco and Martin Johnston.
\newblock {\em {Non-binary error control coding for wireless communication and data storage}}.
\newblock John Wiley \& Sons, 2008.

\bibitem{moon2020error}
Todd~K Moon.
\newblock {\em {Error correction coding: mathematical methods and algorithms}}.
\newblock John Wiley \& Sons, 2020.

\bibitem{pecorari2025high}
Laura Pecorari, Sven Jandura, Gavin~K Brennen, and Guido Pupillo.
\newblock {High-rate quantum LDPC codes for long-range-connected neutral atom registers}.
\newblock {\em Nature Communications}, 16(1):1111, 2025.

\bibitem{babar2015fifteen}
Zunaira Babar, Panagiotis Botsinis, Dimitrios Alanis, Soon~Xin Ng, and Lajos Hanzo.
\newblock {Fifteen years of quantum LDPC coding and improved decoding strategies}.
\newblock {\em IEEE Access}, 3:2492--2519, 2015.

\bibitem{rengaswamy2020optimality}
Narayanan Rengaswamy, Robert Calderbank, Michael Newman, and Henry~D Pfister.
\newblock {On optimality of CSS codes for transversal T}.
\newblock {\em IEEE Journal on Selected Areas in Information Theory}, 1(2):499--514, 2020.

\bibitem{calderbank1996good}
A~Robert Calderbank and Peter~W Shor.
\newblock {Good quantum error-correcting codes exist}.
\newblock {\em Physical Review A}, 54(2):1098, 1996.

\bibitem{fossorier1999reduced}
Marc~PC Fossorier, Miodrag Mihaljevic, and Hideki Imai.
\newblock {Reduced complexity iterative decoding of low-density parity check codes based on belief propagation}.
\newblock {\em IEEE Transactions on communications}, 47(5):673--680, 1999.

\bibitem{ferraz2023unlocking}
Oscar Ferraz, Yann Falevoz, Vitor Silva, and Gabriel Falcao.
\newblock {Unlocking the Potential of LDPC Decoders with PiM Acceleration}.
\newblock In {\em 2023 57th Asilomar Conference on Signals, Systems, and Computers}, pages 1579--1583. IEEE, 2023.

\end{thebibliography}

\end{document}